\documentclass[pdflatex,sn-mathphys-ay]{sn-jnl}

\usepackage{graphicx}%
\usepackage{multirow}%
\usepackage{amsmath,amssymb,amsfonts}%
\usepackage{amsthm}%
\usepackage{mathrsfs}%
\usepackage[title]{appendix}%
\usepackage{xcolor}%
\usepackage{textcomp}%
\usepackage{manyfoot}%
\usepackage{booktabs}%
\usepackage{algorithm}%
\usepackage{algorithmicx}%
\usepackage{algpseudocode}%
\usepackage{listings}%

\begin{document}

\renewcommand{\thetable}{\Roman{table}}

\title[Article Title]{Comparative Analysis of Data Preprocessing Methods, Feature Selection Techniques and Machine Learning Models for Improved Classification and Regression Performance on Imbalanced Genetic Data}


\author*[1]{\fnm{Arshmeet} \sur{Kaur}}\email{Arka7783@stu.evc.edu}

\author[2]{\fnm{Morteza} \sur{Sarmadi}}\email{msarmadi@mit.edu}

\affil*[1]{\orgdiv{Biology Student}, \orgname{Evergreen Valley College}, \orgaddress{\street{3095 Yerba Buena Rd}, \city{San Jose}, \postcode{95135}, \state{CA}, \country{U.S.A}}}

\affil[2]{\orgdiv{Research Scientist}, \orgname{Gilead Sciences}, \orgaddress{\street{333 Lakeside Dr}, \city{Foster City}, \postcode{94404}, \state{CA}, \country{U.S.A}}}


\abstract{Rapid advancements in genome sequencing have led to the collection of vast amounts of genomics data. Researchers may be interested in using machine learning models on such data to predict the pathogenicity or clinical significance of a genetic mutation. However, many genetic datasets contain imbalanced target variables that pose challenges to machine learning models: observations are skewed/imbalanced in regression tasks or class-imbalanced in classification tasks. Genetic datasets are also often high-cardinal and contain skewed predictor variables, which poses further challenges. We aimed to investigate the effects of data preprocessing, feature selection techniques, and model selection on the performance of models trained on these datasets. We measured performance with 5-fold cross-validation and compared averaged r-squared and accuracy metrics across different combinations of techniques. We found that outliers/skew in predictor or target variables did not pose a challenge to regression models. We also found that class-imbalanced target variables and skewed predictors had little to no impact on classification performance. Random forest was the best model to use for imbalanced regression tasks. While our study uses a genetic dataset as an example of a real-world application, our findings can be generalized to any similar datasets.}

\maketitle

\section{Introduction}\label{sec1}

When dealing specifically with predicting the pathogenicity/clinical significance of a mutation, we face two problems: either imbalanced regression or class-imbalanced classification. Our study aims to find methods that lead to the best performance of imbalanced regression and class-imbalanced classification tasks for models trained on data with the two characteristics mentioned above. While we are using a genetic dataset, our results are applicable to any datasets that share those characteristics of high-cardinality and skewed predictors. 

The first problem, imbalanced regression, occurs when machine learning models aim to predict a continuous, skewed target variable. For example, consider CADD\_PHRED, a score ranging from 0 (benign) to 1 (most deleterious) \citep{niroula2019good}. Most missense mutations (>70\%) are mildly deleterious \citep{kryukov2007most}. Additionally, medical researchers studying disease might be more likely to focus on deleterious mutations. If they obtain data from a source focusing on the clinical/disease significance of genetic variants, such as Clinvar, observations may be skewed toward deleterious mutations. Thus, we would expect CADD\_PHRED, in the dataset used here, to be skewed heavily towards deleterious mutations.

The imbalanced regression problem poses two main challenges \citep{ribeiro2020imbalanced}: 1) normal distributions are a common assumption of machine learning models and earlier utility-based regression. 2) Attempts to optimize model performance often result in severe bias. Many real-world target variables must be modeled as continuous variables, as categories do not provide enough information. Additionally, the majority of the research on imbalanced target variables focuses on classification tasks. To deal with datasets containing skewed predictors and targets, past research has attempted to use preprocessing methods to address skewed predictor variables \citep{branco2019study}. Another possible solution might be different feature selection techniques or model choices; while there have been many studies focusing on feature selection and dimensionality reduction for class-imbalanced classification problems \citep{Kryukov2007MostRM}, \citep{Maldonado2014FeatureSF}, \citep{Khoshgoftaar2010ANF}, \citep{Pant2015ASO}, as pointed out by \citep{branco2019study}, there is a research gap in studying the data-intrinsic characteristics for imbalanced regression tasks. Following suit, there have been many studies focusing on the machine learning models that work best on class-imbalanced classification problems \citep{luo2019logistic}, \citep{Esteves2020TechniquesTD}, \citep{mirza2016efficient}, but to our knowledge, there are few for imbalanced regression tasks. The first part of our study aims to supplement this sparse research by assessing which data preprocessing techniques (log or power transforms), feature selection techniques (univariate and embedded), and model choice (between decision trees, K-nearest neighbors, RANSAC, random forest, and support vector regressor) deal with our high-cardinality data with both skewed predictor and target variables. In other words, we aim to determine which combination of these three factors leads to the best performance of the regression task (predicting CADD\_PHRED).

The second problem is class-imbalanced classification. Pathogenicity scores can be categorized as “benign” or “damaging” based on a cutoff (for example, those specified on Ensembl \citep{ensembl2014pathogenicity}). In this study, we look at SIFT scores, which determine whether a genetic variant will cause a change in protein function, with 0 being the most deleterious and 1 being tolerated \citep{sim2012sift}. In the dataset used, the scores are categorized as deleterious, tolerated, tolerated low confidence, and deleterious low confidence. PolyPhen scores (also 0 to 1) represent the probability of a given mutation being deleterious to protein structure and function: They can be sorted into benign, probably damaging, possibly damaging, and unknown. PolyPhen and SIFT represent the effect of a mutation on protein structure, and only approximately 20 percent of non-synonymous SNPs (single nucleotide polymorphisms) are deleterious to protein structure \citep{sunyaev2001prediction}. Thus, we would expect there to be more observations in the “tolerated” classes, leading to class imbalance.

The class-imbalance problem occurs when there is a disproportionate amount of observations in each class of a classification target, leading machine learning algorithms to treat the less common class values as noise \citep{abd2013review}. It is important to address this problem because class imbalance is often observed in biomedical, medical diagnostic, and bioinformatic data, as well as in other fields like geology or finance. Using different feature selection techniques is one way to address the problem; for example, researchers have developed an embedded feature selection approach derived from the SVM model \citep{nguwi}. There has also been some research looking at the effect of model choice for class-imbalanced datasets with skewed predictor variables, but not many models have been tested; one study found that naive Bayes Classifier outperformed Probabilistic Neural Network \citep{Shahadat2015AnEA}. Our research will focus on which data preprocessing (log and power transforms), feature selection techniques (univariate feature vs embedded selection), and model choice (between decision trees, random forest, K-nearest neighbors, Naive Bayes Classifier, Support Vector Classifier) deal with our high-cardinality data with both skewed predictor and imbalanced target variables best. In other words, again, we seek to find which combination will lead to the best performance of the classification tasks.

To summarize, this research aims to look at a variety of data transformation techniques, feature selection techniques, and model choices to deal with skewed predictor variables and either skewed or class-imbalanced target variables. Our results apply to datasets similar to ours. Throughout the paper, the sections will be split to separately address the regression task and classification tasks. 

\section{Methods}\label{sec2}

\subsection{Data Preprocessing and Creation of Datasets For Model Training}
\begin{figure}[h]
\centering
\includegraphics[width=0.6\textwidth]{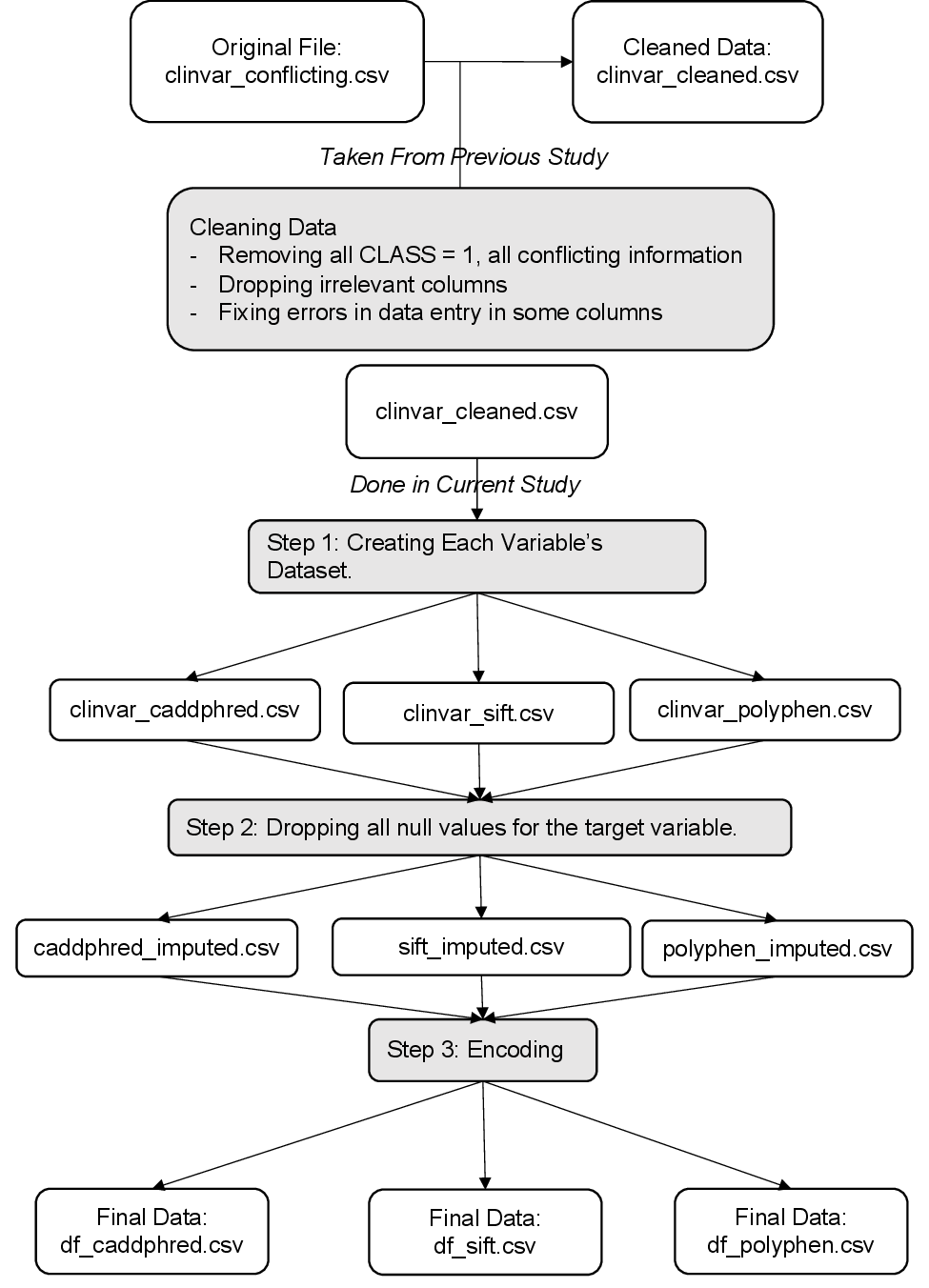}
\caption{Workflow of Data Preprocessing: The dataset used in this research, clinvar\_cleaned.csv was obtained from the authors of \citep{Kaur2024PredictingLI}, who cleaned an original dataset from \citep{kevin_arvai_2020}. The process of creating the cleaned dataset (removing conflicting information, dropping irrelevant\/low\-variance variables, and fixing errors in data entry) is described in sections 2.1 to 2.3 of \citep{Kaur2024PredictingLI}. The rest of this figure details the datasets created in this study.}\label{fig1}
\end{figure}

We obtained a cleaned dataset from \citep{Kaur2024PredictingLI}, originally from \citep{kevin_arvai_2020} (see Fig.1). In step one, we made three copies of the cleaned dataset to create three different datasets for each of the target variables in this paper: CADD\_PHRED, SIFT and PolyPhen. The cleaned dataset contained LoFtool, BLOSUM62, CADD\_PHRED, PolyPhen, and SIFT scores. The combinations of data transformations, feature selection techniques, and model choices that lead to the best performance for LoFtool and BLOSUM62 have already been separately studied. Additionally, BLOSUM62 is more of a probability-of-substitution score than a pathogenicity score \citep{eddy2004did}. Therefore, we deemed both variables irrelevant to this paper. We dropped the “score” variables other than the target variable from each respective dataset. For example, to create the PolyPhen dataset (clinvar\_polyphen.csv), we dropped BLOSUM62, LoFtool, CADD\_PHRED, and SIFT. In step two, we dropped all null values for each target variable, reducing the number of entries in each dataset from 39693 to 39454, 18246, and 18229 for CADD\_PHRED, SIFT, and PolyPhen respectively. In step three, we performed encoding for categorical variables of all datasets created in step two. For all datasets, we used regularized target encoding, with CADD\_PHRED, SIFT, or PolyPhen as the “target.” Regularized target encoding has been shown to outperform other common methods such as leaf or one-hot encoding for high-cardinality data \citep{pargent2022regularized}. For SIFT (categories: tolerated low confidence, tolerated, deleterious low confidence and deleterious and PolyPhen (categories: benign, probably damaging, possibly damaging), ordinal encoding was used first to assign numerical values 1.0, 2.0… in order of increasing pathogenicity and then regularized target encoding was applied. 

\begin{figure}[h]
\centering
\includegraphics[width=0.9\textwidth]{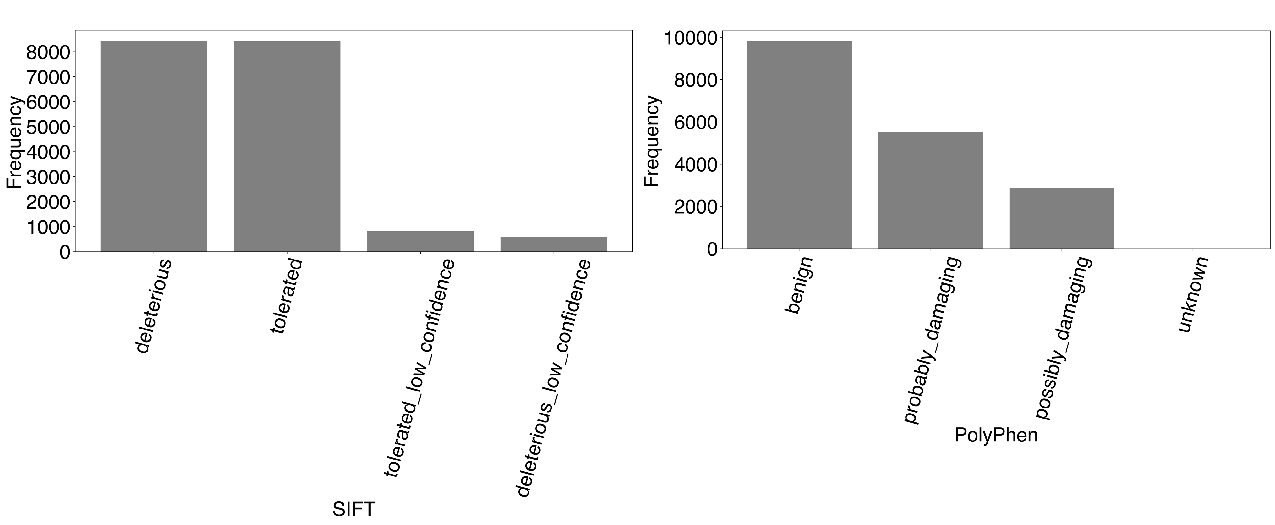}
\caption{SIFT and Polyphen Class Distribution: In this figure, we visualized SIFT and PolyPhen class distributions in order to assess if the class-imbalance problem was present in each dataset \citep{guo}. 
}\label{fig2}
\end{figure}

\begin{figure}[h]
\centering
\includegraphics[width=0.5\textwidth]{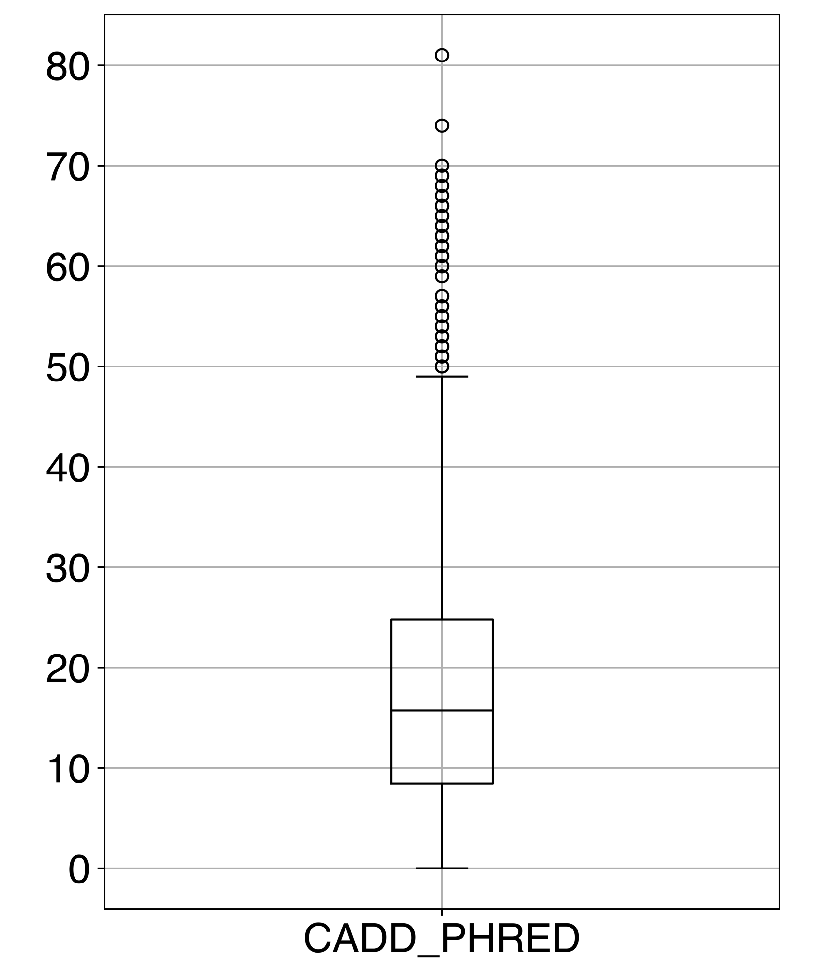}
\caption{CADD\_PHRED's Heavy Right Skew: The presence of outliers and heavy right skew in CADD\_PHRED indicates that models training on this data have to deal with the imbalanced regression problem, which is explained in greater detail in the introduction section.}
\label{fig3}
\end{figure}
Fig. 2 illustrates the class distribution of the SIFT and PolyPhen variables in their datasets. Because there were only two observations in the “unknown” category, we deemed that class irrelevant. A decent amount of observations were present in the remaining classes. However, for the SIFT variable in the SIFT dataset, the two “low\_confidence” categories had fewer observations than the tolerated and deleterious categories did. We also observed outliers and heavy right skew in the CADD\_PHRED variable (see Fig. 3). Finally, as visualized in Fig. 4, across all datasets, six predictor variables were consistently heavily skewed. 
\begin{figure}[h]
\centering
\includegraphics[width=1\textwidth]{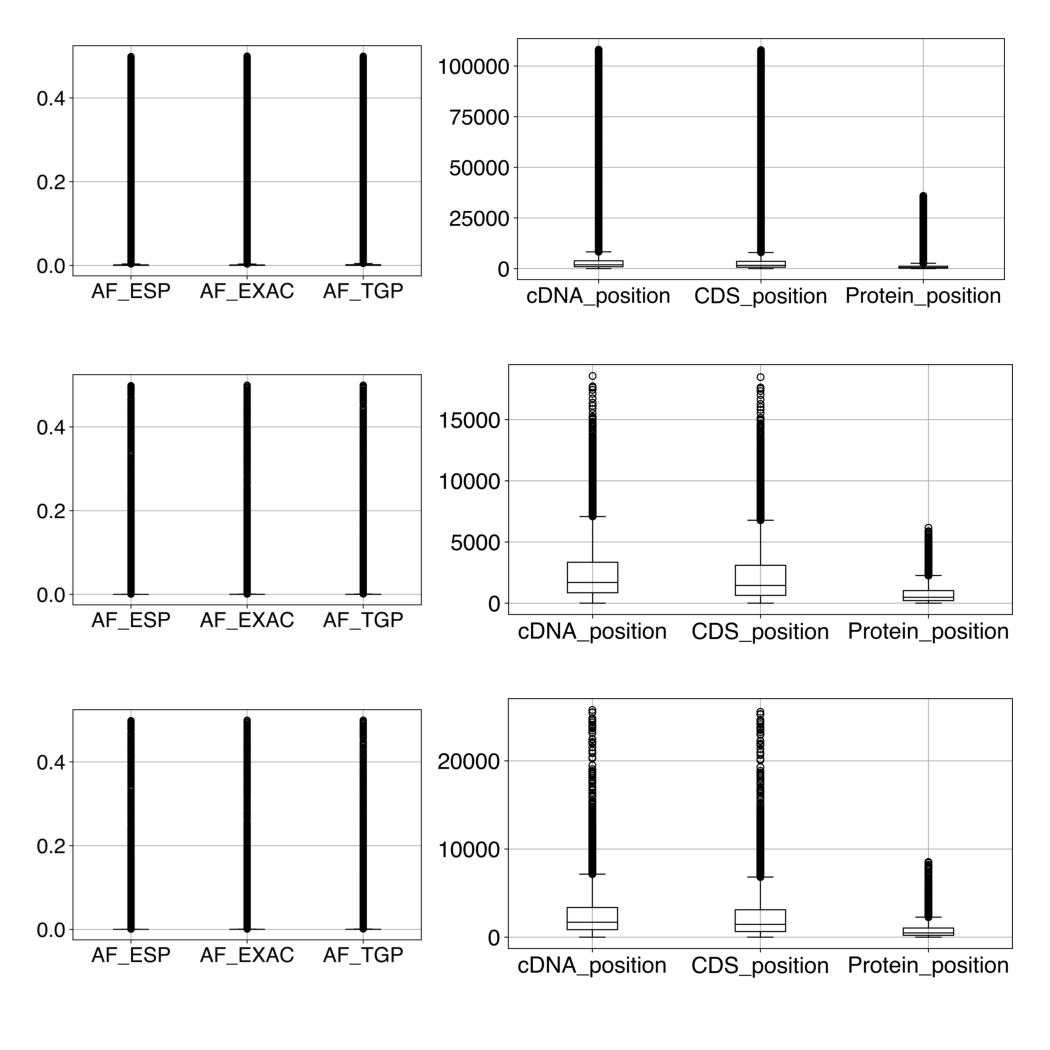}
\caption{Comparing Continuous Variables Across Datasets: We found six variables (AF\_ESP, AF\_EXAC,  AF\_TGP, cDNA\_position, CDS\_position, Protein\_position) that were heavily skewed in all three datasets (df\_polyphen.csv, df\_sift.csv, and df\_caddphred.csv for the top, middle and bottom panels respectively).}
\label{fig4}
\end{figure}

\begin{table}[h]
\caption{Datasets Created for Regression Task Model Training}\label{tab:datasets}
\begin{tabular}{@{}llll@{}}
\toprule
\textbf{Set} & \textbf{Name of Dataset} & \textbf{Handling CADD\_PHRED} & \textbf{Handling Skewed Predictors} \\
\midrule
1 & df\_caddphred & Not applicable. & Not applicable. \\
  & df\_caddphred\_yj & Yeo-Johnson Transformed & Yeo-Johnson Transformed. \\
  & df\_caddphred\_log & Logarithm Transformed & Logarithm Transformed \\
\midrule
2 & caddphred\_no\_outliers\_encoded & Dropped Outliers & Not applicable. \\
  & caddphred\_no\_outliers\_encoded\_yj & Dropped Outliers & Yeo-Johnson Transformed. \\
  & caddphred\_no\_outliers\_encoded\_log & Dropped Outliers & Logarithm Transformed \\
\bottomrule
\end{tabular}
\footnotetext{We created two sets, each with unique ways of handling the skewed predictor and target variables. We will make within-set and between-set comparisons to compare different methods of transformation and determine whether skewed predictor variables affect model performance. Set 1 contains df\_caddphred (the final dataset for the CADD\_PHRED target variable) and its transforms. All outliers (1.5 times the IQR above or below the first or third quartile) were dropped from df\_caddphred to create caddphred\_no\_outliers\_encoded. Set 2 contains caddphred\_no\_outliers\_encoded and its transforms.}
\end{table}

\begin{table}[h]
\caption{Datasets Created for Classification Tasks Model Training}\label{tab:classification_datasets}
\begin{tabular}{lll}
\toprule
\textbf{Set} & \textbf{Dataset} & \textbf{Handling Skewed Predictors} \\
\midrule
3 & df\_polyphen & Not applicable. \\
  & df\_polyphen\_yj & Yeo-Johnson Transformed. \\
  & df\_polyphen\_log & Logarithm Transformed \\
\midrule
4 & df\_sift & Not applicable. \\
  & df\_sift\_yj & Yeo-Johnson Transformed. \\
  & df\_sift\_log & Logarithm Transformed \\
\bottomrule
\end{tabular}
\footnotetext{We created Set 3 from df\_polyphen and Set 4 from df\_sift (see Fig. 1 for more information on these datasets). We transformed skewed predictor variables in both sets. Nothing was done to modify the class-imbalanced categorical target variables in both sets. We will primarily be comparing the sets to assess the effect of skewed predictor variables on classification performance.}
\end{table}

To deal with the skewed predictor variables and skewed CADD\_PHRED target variable, we will use logarithm transformation \citep{curran2018explorations}, and Yeo-Johnson transformation (similar to Box-Cox) \citep{weisberg2001yeo}, \citep{yeo2000new}. Researchers have questioned the validity of the logarithm transform in biomedical research and data analysis \citep{changyong2014log}, \citep{feng2013log}, \citep{keene1995log}. Thus, it could be insightful to compare model performance on data that has been transformed in different ways. We created a multitude of training datasets from each target variable’s dataset (refer to Tables 1 and 2). We have a total of six datasets for the regression task and six for the classification task. 

\subsection{Training and Testing Models}
\begin{figure}[h]
\centering
\includegraphics[width=1\textwidth]{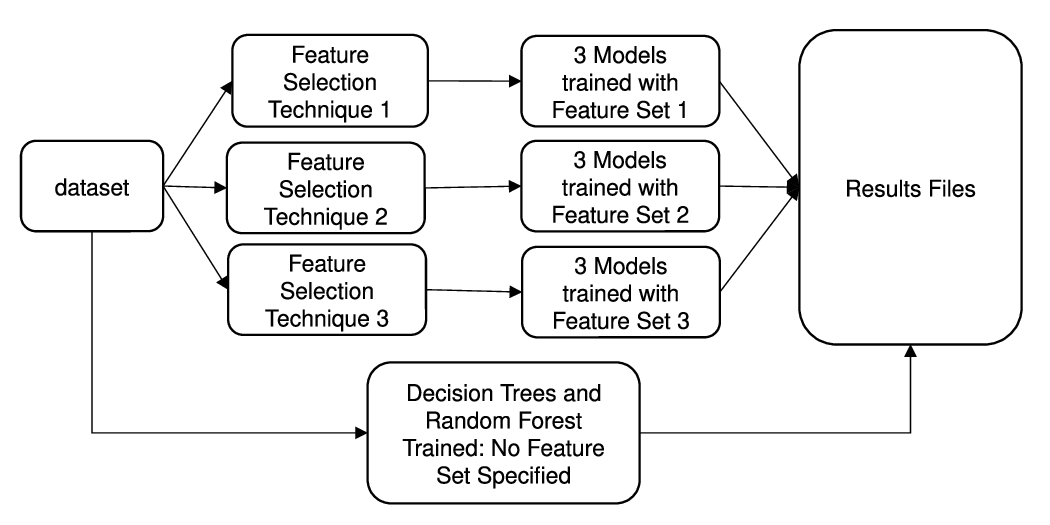}
\caption{Examples of Combinations of Dataset, Feature Selection Techniques, and Model Choices: Each of the twelve datasets from Sets 1 through 4 (see Tables 1 and 2) produced eleven unique combinations of the dataset used, feature selection technique applied, and model trained. Three feature sets were created for each dataset using different univariate feature selection techniques, and three models were trained for each feature set. In addition, for every dataset, two more models, random forest and decision trees were trained with no specified feature set, allowing the models to use their embedded feature selection techniques. In total, there were twelve datasets in Sets 1-4 and eleven combinations for each, meaning we tested one hundred and thirty-two total combinations. Results for the regressors were saved to one file and classifiers to another.}
\label{fig5}
\end{figure}

We planned to compare the performance of different feature selection techniques. Theoretically, univariate feature selection techniques should yield worse performance than other techniques in bioinformatics research \citep{ni2012review}; to the contrary, some studies show that in research, univariate methods can yield better results. However, this may be chalked up to having a limited sample size of studies. Therefore, we aim to compare model performance between different feature selection techniques for classification and regression tasks. For the classification task (predicting SIFT and PolyPhen), we will use f-classification, chi-squared classification, and mutual-info classification (all univariate feature selection techniques from sklearn) as well as the embedded feature selection of decision trees and random forest classifiers. For the regression task (predicting CADD\_PHRED), we will use f-regression, r-regression, and mutual-info regression as well as the embedded feature selection of decision trees and random forest regressors. For all univariate feature selection techniques applied, we used SelectKBest from sklearn in order to make feature sets of ten. We also chose to compare performances between different machine learning models.  For the regression task, we chose to compare between decision trees, K-nearest neighbors, RANSAC (robust to outliers \citep{zuliani2009ransac},  \citep{derpanis2010overview}), random forest, and support vector regressor. For the classification task, we chose to compare decision trees, random forests, K-nearest neighbors, Naive Bayes Classifier, and Support Vector Classifier. We used 5-fold cross-validation (to reduce overfitting concerns \citep{charilaou2022machine},  \citep{montesinos2022overfitting}) and measured the averaged r-squared or accuracy for each combination, which will show us the goodness-of-fit of all models. 

\subsection{Analysis of Results}
To analyze the results (r-squared and accuracy for all combinations), we used robustlmm, an extension of the lme4 R package for linear mixed effects modeling \citep{koller2016robustlmm}, \citep{bates2014fitting}. Linear mixed effects models provide an appropriate regression for complex datasets with multiple variables where observations are “clustered.” The results file contains three characteristics: the model, dataset, and feature selection technique used. It also includes the r-squared associated with each unique combination of these three. Some of the observations share two of these three characteristics, making them more closely related to each other than to other observations. Observations in this data could share both dataset and feature selection techniques, for example. We will refer to the variable that forms these clusters as a “grouping variable.” In the case of our dataset, the grouping variable was a combined variable of data plus feature selection, data plus model choice, and feature selection plus model choice. Mixed models can contain fixed effects, variables that do not vary across different values of the grouping variable, and random effects, which do vary. The random effects account for variability in the response variable (r-squared) that cannot be accounted for by the fixed effects. For example, the fixed effects were the combined grouping variables (data/feature selection, data/model, and feature selection/model) and the corresponding random effects were model, feature selection, or data, respectively. For example, when analyzing the effect that using a certain dataset (random effect) had on the response variable (r-squared), feature selection/model was the grouping variable and a fixed effect. 

We used robust linear mixed-effects modeling from the robustlmm package in R, which can handle contamination and outliers that regular linear mixed-effect models may not be able to \citep{koller2016robustlmm},. For the best outcome, we tested the assumptions of linear mixed-effect models: linearity, homoscedasticity (homogeneity of variances), and normality of the residuals. To test these, we loosely followed the guidelines from the University of Illinois at Chicago \citep{palmeri2016testing}. To test the linearity assumption, we plotted residuals against fitted values and looked for random scatter around y = 0. Homoscedasticity could be assessed from the same plot: essentially, we were looking for random scatter, no “fanning out.” Lastly, we tested the normality of the residuals using a histogram of residual values (looking for normal distribution) and a Q-Q plot (looking for points following the qline generated by R closely). Research indicates that linear mixed models are quite robust to these distributional assumptions being violated \citep{schielzeth2020robustness}; to be careful, we will be assessing the Intraclass Correlation Coefficient \citep{koo2016guideline} and Marginal and Conditional r-squared values to see the goodness–of–fit of the model.

\section{Results}\label{sec3}
Our main question had three components: which data transformations, feature selection techniques, and machine learning models lead to increased r-squared (for regression tasks) or accuracy (for classification tasks)? Thus, we split our results accordingly. The best-performing combination for the regression task was random forest, using its own embedded feature selection, trained on df\_caddphred with an r-squared value of 0.64. The best-performing combination for the classification task was the SVC classifier, trained on df\_sift\_log, using chi-squared feature selection and it achieved an accuracy of 0.66.

\subsection{Between-Set and Within-Set Comparisons of Data Transformations Effects on Regression Task Performance}

In the output of mixed-effects models, the intercept term, marked as “(Intercept),” represents the baseline value of the response variable. For example, when we use the intercept dataset caddphred\_no\_outliers\_encoded in Table 3, we expect a value of 0.39 for the response variable. All other values in the table are “adjustments” of this baseline value; for example, we see a 0.05(0.00-0.09) increase in the response variable when the caddphred\_no\_outliers\_encoded\_log dataset is used. In Tables 3 to 6, the “response” variable is the r-squared value measuring the goodness-of-fit for the machine learning models trained and tested. This is not to be confused with the marginal and conditional R-squared values stated at the bottom of each table; these values refer to the goodness-of-fit of the linear mixed-effects regression model used to analyze results files (see Fig. 4 for more information on the results files). Marginal R-squared values show us the percentage of variance explained by fixed effects whereas conditional R-squared values show variance explained by both fixed and random effects \citep{nakagawa2013general}. Simply put, the random effects are the variables you see in each table. In Table 3, the data variable is the random effect and the fixed effect is a combined model choice and feature selection variable. More detail on this is given in the methods section. We will use a significance cutoff of p$<$0.05.

\begin{table}[h]
\caption{R-Squared Adjustment Values Based On Dataset (Sets 1 and 2)}\label{tab:rsquared_adjustment}
\begin{tabular}{@{}lllll@{}}
\toprule
\textbf{Set} & \textbf{Predictors} & \textbf{Estimates} & \textbf{CI} & \textbf{p} \\
\midrule
2 & (Intercept) [caddphred\_no\_outliers\_encoded] & 0.39 & 0.24 – 0.54 & $<$0.001 \\
  & data [caddphred\_no\_outliers\_encoded\_log] & 0.05 & 0.00 – 0.09 & 0.041 \\
  & data [caddphred\_no\_outliers\_encoded\_yj] & 0.03 & -0.02 – 0.07 & 0.203 \\
\midrule
1 & data [df\_caddphred] & 0.03 & -0.02 – 0.08 & 0.196 \\
  & data [df\_caddphred\_log] & -0.35 & -0.40 – -0.31 & $<$0.001 \\
  & data [df\_caddphred\_yj] & -0.01 & -0.05 – 0.04 & 0.690 \\
\midrule
\textbf{Random Effects} & & & & \\
\midrule
 & $\sigma^2$ & 0.00 & & \\
 & $\tau_{00}$ grouping\_variable & 0.06 & & \\
 & ICC & 0.95 & & \\
 & N grouping\_variable & 11 & & \\
\midrule
\textbf{Observations} & & & & \\
\midrule
 & & & & \\
 & \textbf{Marginal R2 / Conditional R2} & 0.245 / 0.965 & & \\
\bottomrule
\end{tabular}
\footnotetext{R-squared adjustment values (estimates) based on predictor variables (dataset used in this case), confidence intervals, and p-values. In this case, the low marginal R-squared and high conditional R-squared of 0.97 indicate that a large amount of the variance in the response variable (machine learning models’ r-squared values) is explained by the dataset used. An ICC of 0.95 indicates the model’s reliability \citep{bobak2018estimation}}.
\end{table}

In Table 3, we conducted two comparisons between Set 1 and Set 2. Firstly, we compared the r-squared adjustment values for df\_caddphred (outliers present) and caddphred\_no\_outliers\_encoded (outliers dropped) to see if dropping outliers (thus normalizing skew of the target variable) had any impact on model performance. We found no significant difference between the two datasets, indicating that the outliers and skewed target variables had no impact on model performance. 

We also compared the caddphred\_no\_outliers\_encoded\_log and df\_caddphred\_log datasets. The only difference between these datasets was that the former dropped outliers and the latter transformed them. Both dealt with skewed predictor variables the same way, by log transforming them. Comparing these datasets allows us to see the effect of log-transforming outliers versus dropping them when predictor variables are normalized using the log transformation; using the caddphred\_no\_outliers\_encoded\_log dataset led to a marginal 0.05 (0.00-0.09) increase in r-squared while using the df\_caddphred\_log dataset led to a -0.35 (-0.05-0.04) decrease in r-squared. To summarize, when predictors are log-transformed, dropping outliers in the target variable led to better model performance than log-transforming them.  However, dropping outliers only caused a marginal increase as compared to the baseline.

\begin{table}[h]
\caption{R-Squared Adjustment Values Based On Dataset (Set 1)}\label{tab:rsquared_adjustment_set1}
\begin{tabular}{@{}lllll@{}}
\toprule
\textbf{Predictors} & \textbf{Estimates} & \textbf{CI} & \textbf{p} \\
\midrule
(Intercept) [df\_caddphred] & 0.42 & 0.26 – 0.57 & $<$0.001 \\
data [df\_caddphred\_log] & -0.38 & -0.44 – -0.32 & $<$0.001 \\
data [df\_caddphred\_yj] & -0.04 & -0.10 – 0.02 & 0.190 \\
\midrule
\textbf{Random Effects} & & & \\
\midrule
$\sigma^2$ & 0.00 & & \\
$\tau_{00}$ grouping\_variable & 0.06 & & \\
ICC & 0.93 & & \\
N grouping\_variable & 11 & & \\
\midrule
\textbf{Observations} & & & \\
\midrule
 & & & \\
\textbf{Marginal R2 / Conditional R2} & 0.309 / 0.951 & & \\
\bottomrule
\end{tabular}
\footnotetext{A marginal R-squared of 0.31 with a conditional R-squared of 0.95 for the regression indicates that a relatively large percentage of the variance is explained by random effects (levels of the data variable). The ICC showed reliability, \citep{bobak2018estimation}}.
\end{table}

In addition to between-set comparisons, we also carried out within-set comparisons. Several variables were skewed (Fig. 3) in the original CADD\_PHRED dataset. Within Set 1, two of the datasets transform and normalize both the predictor and target variables using the log and Yeo-Johnson transforms. If we compare datasets within Set 1, we can see if the transformation was worthwhile and whether the log or Yeo-Johnson transformation was better. As we can see in Table 4, there was no significant difference between the untransformed, original CADD\_PHRED dataset (df\_caddphred) and the Yeo-Johnson transform (df\_caddphred\_yj). Log transforming predictor and target variables (df\_caddphred\_log) significantly decreased model performance by -0.38 (-0.44– -0.32). Therefore, we conclude that transforming both predictor and target variables with either the log-transform or Yeo-Johnson transform brought about no benefit.

\begin{table}[h]
\caption{R-Squared Adjustment Values Based On Dataset (Set 2)}\label{tab:r_squared}
\centering
\begin{tabular}{@{}llll@{}}
\toprule
\textbf{Predictors} & \textbf{Estimates} & \textbf{CI} & \textbf{p} \\
\midrule
(Intercept) [caddphred\_no\_outliers\_encoded] & 0.40 & 0.25 -- 0.54 & $<$0.001 \\
data [caddphred\_no\_outliers\_encoded\_log] & 0.05 & -0.00 -- 0.09 & 0.056 \\
data [caddphred\_no\_outliers\_encoded\_yj] & 0.03 & -0.02 -- 0.08 & 0.211 \\
\midrule
\textbf{Random Effects} & & & \\
$\sigma^2$ & 0.00 & & \\
$\tau_{00}$ grouping\_variable & 0.06 & & \\
\midrule
\textbf{ICC} & & & \\
ICC & 0.95 & & \\
\midrule
\textbf{N grouping\_variable} & & & \\
N & 11 & & \\
\midrule
\textbf{Observations} & & & \\
Observations & 33 & & \\
\midrule
\textbf{Marginal R2 / Conditional R2} & 0.006 / 0.951 & & \\
\bottomrule
\end{tabular}
\footnotetext{ A near-nonexistent marginal R-squared with a conditional R-squared of 0.95 for the regression indicates that most of the variance in the machine learning models' r-squared values is explained by random effects (levels of the data variable). An ICC of 0.95 indicates the model’s reliability~\citep{bobak2018estimation}.}
\end{table}

Within Set 2, all sets have a normalized target variable (outliers were dropped in the target variable). Two of the sets have log or Yeo-Johnson transformed the skewed predictor variables. Comparing datasets within Set 2 allows us to examine whether the skew of predictor variables is important to model performance once the skew of the target variable has already been normalized. As seen in Table 5, there was no significant change in performance caused by either the log or Yeo-Johnson transforms. This seems to indicate that the skew of the predictor variables does not make a difference when the skew of the target variable is normalized. 

\subsubsection{Comparing Different Model Choice}

\begin{table}[h]
\caption{R-squared Adjustments For Different Model Choices (Sets 1 and 2)}\label{tab:model_choices}
\begin{tabular}{@{}lllll@{}}
\toprule
\textbf{Predictors} & \textbf{Estimates} & \textbf{CI} & \textbf{p} \\
\midrule
(Intercept) [dt\_regression] & 0.16 & -0.07 – 0.39 & 0.163 \\
model [knn\_regression] & 0.19 & -0.07 – 0.45 & 0.150 \\
model [ransac\_regression] & 0.07 & -0.19 – 0.33 & 0.584 \\
model [rf\_regression] & 0.41 & 0.33 – 0.50 & $<$0.001 \\
model [svr\_regression] & 0.26 & -0.00 – 0.52 & 0.051 \\
\midrule
\textbf{Random Effects} & & & \\
\midrule
$\sigma^2$ & 0.00 & & \\
$\tau_{00}$ grouping\_variable & 0.07 & & \\
ICC & 0.93 & & \\
N grouping\_variable & 24 & & \\
\midrule
\textbf{Observations} & & & \\
\midrule
 & & & \\
\textbf{Marginal R2 / Conditional R2} & 0.147 / 0.944 & & \\
\bottomrule
\end{tabular}
\footnotetext{The marginal and conditional R-squared values show that the majority of variance is explained by random effects (levels of the data variable) and the ICC shows the regression is reliable \citep{bobak2018estimation}}
\end{table}

For both sets, Random forest was the most successful model, causing an increase of 0.41 (0.33-0.50). The others may look somewhat close in effect size, but none are statistically significant. SVR was in second with a 0.26 (-0.00-0.52) but just marginally did not make the statistical significance cut-off (p = 0.051). 

\subsection{Classification Task Performance}

The classification tasks pose a less complex task as the data is in the outcome variables (SIFT and PolyPhen). However, as mentioned in the methods section, the SIFT dataset has a class imbalance whereas the PolyPhen variable does not. We will observe whether there is a difference in which combination of data transformation techniques, feature selection techniques, and model choice gives the best performance and whether this differs between the two variables’ datasets (Sets 3 and 4 from Table 2).

\subsubsection{Comparing different Data Transformation Techniques}

\begin{table}[h]
\caption{Accuracy Adjustments Based on Dataset (Sets 3 and 4)}\label{tab:accuracy_adjustments}
\centering
\begin{tabular}{@{}llll@{}}
\toprule
\textbf{Predictors} & \textbf{Estimates} & \textbf{CI} & \textbf{p} \\
\midrule
(Intercept) [df\_polyphen] & 0.56 & 0.52 -- 0.61 & $<$0.001 \\
data [df\_polyphen\_log] & 0.00 & -0.02 -- 0.02 & 0.710 \\
data [df\_polyphen\_yj] & 0.00 & -0.02 -- 0.03 & 0.659 \\
data [df\_sift] & 0.01 & -0.01 -- 0.03 & 0.209 \\
data [df\_sift\_log] & 0.02 & -0.00 -- 0.04 & 0.056 \\
data [df\_sift\_yj] & 0.02 & -0.00 -- 0.04 & 0.123 \\
\midrule
\textbf{Random Effects} & & & \\
$\sigma^2$ & 0.00 & & \\
$\tau_{00}$ combined\_factor & 0.00 & & \\
\midrule
\textbf{ICC} & & & \\
ICC & 0.88 & & \\
\midrule
\textbf{N combined\_factor} & & & \\
N & 11 & & \\
\midrule
\textbf{Observations} & & & \\
Observations & 66 & & \\
\midrule
\textbf{Marginal R2 / Conditional R2} & 0.011 / 0.886 & & \\
\bottomrule
\end{tabular}
\footnotetext{Shown in the table are the dataset predictors (which explain most of the variance in the response variable) from Sets 3 and 4. The ICC is reliable, though not as high as in previous tables \citep{bobak2018estimation}.}
\end{table}

\begin{table}[h]
\caption{Accuracy Adjustments for Different Datasets (Set 3)}\label{tab:accuracy_adjustments_set3}
\centering
\begin{tabular}{@{}llll@{}}
\toprule
\textbf{Predictors} & \textbf{Estimates} & \textbf{CI} & \textbf{p} \\
\midrule
(Intercept) [df\_polyphen] & 0.56 & 0.51 -- 0.61 & $<$0.001 \\
data [df\_polyphen\_log] & -0.00 & -0.00 -- 0.00 & 0.709 \\
data [df\_polyphen\_yj] & -0.00 & -0.00 -- 0.00 & 0.723 \\
\midrule
\textbf{Random Effects} & & & \\
$\sigma^2$ & 0.00 & & \\
$\tau_{00}$ grouping\_variable & 0.01 & & \\
\midrule
\textbf{ICC} & & & \\
ICC & 1.00 & & \\
\midrule
\textbf{N grouping\_variable} & & & \\
N & 11 & & \\
\midrule
\textbf{Observations} & & & \\
Observations & 33 & & \\
\midrule
\textbf{Marginal R2 / Conditional R2} & 0.000 / 0.998 & & \\
\bottomrule
\end{tabular}
\footnotetext{Similarly to Table 7, this table shows the accuracy adjustment values based on datasets within Set 3. 99.8 percent of the variance in accuracy adjustment values is due to the dataset used, and the ICC is most reliable at 1.00 \citep{bobak2018estimation}.}
\end{table}

\begin{table}[h]
\caption{Accuracy Adjustments for Different Datasets (Set 4)}\label{tab:accuracy_adjustments_set4}
\centering
\begin{tabular}{@{}llll@{}}
\toprule
\textbf{Predictors} & \textbf{Estimates} & \textbf{CI} & \textbf{p} \\
\midrule
(Intercept) [df\_sift] & 0.55 & 0.50 -- 0.60 & $<$0.001 \\
data [df\_sift\_log] & 0.03 & 0.01 -- 0.04 & 0.001 \\
data [df\_sift\_yj] & 0.02 & 0.00 -- 0.04 & 0.012 \\
\midrule
\textbf{Random Effects} & & & \\
$\sigma^2$ & 0.00 & & \\
$\tau_{00}$ grouping\_variable & 0.01 & & \\
\midrule
\textbf{ICC} & & & \\
ICC & 0.95 & & \\
\midrule
\textbf{N grouping\_variable} & & & \\
N & 11 & & \\
\midrule
\textbf{Observations} & & & \\
Observations & 33 & & \\
\midrule
\textbf{Marginal R2 / Conditional R2} & 0.018 / 0.956 & & \\
\bottomrule
\end{tabular}
\footnotetext{A low marginal R-squared with a conditional R-squared of over 0.95 for the regression models indicates that a relatively small percentage of the variance is explained by fixed effects and a relatively large amount is explained by random effects (levels of the data variable). An ICC of over 0.95 indicates the model’s reliability~\citep{bobak2018estimation}.}
\end{table}

We began by comparing df\_sift (class imbalance present) and df\_polyphen (class imbalance not present) to assess how much of a problem class imbalance posed on overall model performance. Interestingly, as we can see in Table 7, there was no significant difference between df\_sift and df\_polyphen on model performance, indicating that the class imbalance actually did not cause an issue in the SIFT dataset. We also ran regression models on Sets 3 and 4 separately (Tables 8 and 9 respectively). In Set 3, we found no significant difference between the transformed datasets and df\_polyphen. In Set 4, we found only marginal benefit: 0.03 (0.01-0.04) increase due to log transforming and 0.02(0.00-0.04) increase due to Yeo-Johnson transformation. This indicates that in imbalanced classification tasks, the skew of the predictor variables has minimal effect on model performance. 

\subsubsection{Comparing Different Model Choice}

\begin{table}[h]
\caption{Accuracy Adjustments Based on Different Model Choices (Sets 3 and 4)}\label{tab:accuracy_model_choices}
\centering
\begin{tabular}{@{}llll@{}}
\toprule
\textbf{Predictors} & \textbf{Estimates} & \textbf{CI} & \textbf{p} \\
\midrule
(Intercept) [dt\_classifier] & 0.52 & 0.46 -- 0.57 & $<$0.001 \\
model [knn\_classifier] & 0.02 & -0.04 -- 0.09 & 0.518 \\
model [nbayes\_classifier] & 0.05 & -0.02 -- 0.12 & 0.155 \\
model [rf\_classifier] & 0.11 & 0.07 -- 0.15 & $<$0.001 \\
model [svc\_classifier] & 0.06 & -0.01 -- 0.13 & 0.072 \\
\midrule
\textbf{Random Effects} & & & \\
$\sigma^2$ & 0.00 & & \\
$\tau_{00}$ grouping\_variable & 0.00 & & \\
\midrule
\textbf{ICC} & & & \\
ICC & 0.77 & & \\
\midrule
\textbf{N grouping\_variable} & & & \\
N & 24 & & \\
\midrule
\textbf{Observations} & & & \\
Observations & 66 & & \\
\midrule
\textbf{Marginal R2 / Conditional R2} & 0.138 / 0.804 & & \\
\bottomrule
\end{tabular}
\footnotetext{Shown are the accuracy adjustment values based on model choice. The ICC is semi-reliable but falls short of other regression models in other tables \citep{bobak2018estimation}. The majority of the variance in accuracy is explained by model choice.}
\end{table}

Random forest performed the best classification for both sets. When the regression was fit with all datasets (Sets 3 and 4), the accuracy increase estimate was 0.11 (0.07-0.15). The SVC classifier was close to meeting the statistical significance cutoff (p = 0.072), but its effect size was almost half, at a 0.06 increase in accuracy.

\subsection{Issues with Data Analysis for Feature Selection Techniques Employed In Both Tasks}

Unfortunately, we were not able to make any conclusions on the highest-performing feature selection techniques for either the regression or classification tasks. We ran linear mixed effects regression on each set (Sets 1, 2, 3, 4) separately and then on multiple sets at once (Sets 1 and 2, Sets 3 and 4), but all attempts resulted in very low marginal and conditional R-squared as well as a non-existent ICC (0.00). We leave the construction of more suitable statistical models to analyze the results for future research. 

\section{Discussion and Conclusion}
We begin with our findings from the imbalanced regression task; we found that dropping outliers had no impact on model performance, indicating that outliers/skew in the target variable may not be causing the issue. Our result is a characterization of the problem; outliers hold information and a model trained only on the “main group” of observations is only generalizable to that group. However, as previously established in the introduction section, imbalanced regression has been shown to pose challenges to model performance \citep{ribeiro2020imbalanced}, so there must be some property of datasets like ours where a skewed target variable does not pose a threat to model performance. \textit{Sylvia et al. }carried out a study where they focused on the optimization of models performing imbalanced regression tasks \citep{silva2022model}. They tested using the SERA (Squared Error Relevance Area) optimization parameter. It places focus on errors of outlier values in a regression problem. They tested models using SERA against standard boosting algorithms and found that SERA models performed better. This could be a helpful jump-off point for future research: perhaps the best-performing combinations of models could be selected from this study and regression could be optimized using SERA. 

We found that the skew of predictor variables does not seem to make a difference when the target variable is already normalized through outliers being dropped. Log transforming both predictor and target variables leads to abysmally reduced performance. When a dataset has predictor variables normalized with the log-transform, it’s better to drop outliers in the target instead of log-transforming the target variable, even though dropping outliers still only leads to a marginal benefit. This follows logically from research questioning the validity of the log transformation, especially when dealing with biological data: \textit{Feng et al} point out that log transformation does not decrease the variability of data and the results of whichever statistical model is being built from the log-transformed information is often not relevant for the original data set \citep{changyong2014log}. Likewise, \textit{Keene }suggests that the transformation should be classified separately from other types of transformations, as it’s crucial to look at the data being transformed beyond just assessing non-normality \citep{keene1995log}. Our study can extend these studies which were focusing on statistical models to machine learning models, acting as a confirmation that log-transformation is not a suitable transform to apply to data in an imbalanced regression task on biomedical data.

Random Forest was the best-performing model for both the imbalanced regression and class-imbalanced classification task. Random forest shows robustness to noise \citep{Breiman2001RandomF}, quickness, flexibility, and a unique approach to mining high dimensional data \citep{Ziegler2014MiningDW}. Random forest works well on datasets with many features, even with a small number of observations. Random forest’s in-built feature selection technique can take rank and relationships between features into account \citep{Nyongesa2020VariableSU}. Essentially, each tree in Random Forest builds multiple trees with different subsets of training data. Out-of-bag samples are left out and later tested. At each tree, the Random Forest also considers a randomly selected subset of features. While testing different features and subsets, it keeps track of the importance of each variable. A ranking of variables is created from the importance scores, allowing the model to use the most influential variables.

We found no significant difference in performance between df\_sift and df\_polyphen, indicating that for some reason, class imbalance did not cause an issue in these datasets. As mentioned in the introduction, however, several studies have shown that this is a well-established problem in machine learning. We took a look at the effect of skewed predictor variables on model performance for the classification tasks. For both sets 3 and 4, transforming skewed predictor variables had only a marginal positive effect on the performance of machine learning models. \textit{Yousefi et al }concluded that skewness has a significant effect on classification accuracy, as data distributions in machine learning are expected to have an approximately symmetric distribution with central tendency \citep{Yousefi2016ClassificationCW}. While some studies have indicated that in certain cases, normalizing the input variables leads to increased accuracy in classification tasks \citep{chittineni2012study}, \textit{Yousefi et al }suggest that in the domain of biomedical data, transformation interferes with the transparency of the classification process. Therefore, instead of data transformation, feature selection or model choice is suggested to optimize model performance. To conclude, we recommend further research to look into optimizing the combinations of model choice, feature selection, and data transformations shown to be the most effective here (Random Forest). We also encourage future research to go beyond the limitations of our study; we could not make any conclusions on best or worst feature selection techniques. 

\section{Competing Interests}
The authors declare no competing interests as no funding, grant or other form of support was provided for this work. 

\section{Data Availability Statement}
The data used in this study can be found at \citep{kevin_arvai_2020}. All code and datasets generated in this study can be found at https://doi.org/10.5281/zenodo.10694637.


\clearpage
\newpage
\bibliography{sn-bibliography}


\end{document}